\documentclass[oldversion]{aa} 
\usepackage{graphicx,subfigure}
\usepackage[varg]{txfonts}
\usepackage{lscape}
\usepackage{longtable}
\usepackage{rotating}
\usepackage{color}
\usepackage[normalem]{ulem}

\begin{document}

   \title{New insights into the quasi-periodic X-Ray Burster GS 0836--429}

   \author{E. Aranzana
          \inst{1}
          \and
          C. S\'anchez-Fern\'andez\inst{2}
          \and
	  E. Kuulkers\inst{2}}

   \institute{Department of Astrophysics/IMAPP, Radboud University, PO Box 9010, 6500 GL Nijmegen, The Netherlands.
   \and ISOC, ESA, European Space Astronomy Centre (ESAC), 
    PO Box 78, 28691 Villanueva de la Ca\~nada (Madrid), Spain.}
   \offprints{\email{e.aranzana@astro.ru.nl}}
           
\date{Accepted \today}
  
\abstract{GS 0836--429 is a neutron star X-ray transient that displays Type-I X-ray bursts. In 2003 and 2004 it experienced two outbursts in X-rays. We present here an analysis of the system bursting properties during these outbursts. We studied the evolution of the 2003--2004 outbursts in soft X-rays using {\it RXTE} (2.5--12\,keV; ASM), and in hard X-rays with {\it INTEGRAL} (17--80\,keV, IBIS/ISGRI). Using data from the JEM-X monitor onboard {\it INTEGRAL} we studied the bursting properties of the source. We detected 61 Type-I X-ray bursts during the 2004 outburst, and confirm that the source displayed a quasi-periodic burst recurrence time of about 2.3 hours. We improve the characterization of the fuel composition, as well as the description of the typical burst durations and fluences. We estimate the average value of $\alpha$ to be $49\pm\,3$. This value together with the observed burst profiles indicate a regime of a mixed He/H runaway triggered by unstable helium ignition. In addition, we report the detection of four series of double bursts, with burst recurrence times of $\leq\,20$ minutes. The secondary bursts are always shorter and less energetic than the primary and typical bursts from the source. The measured recurrence time  in double bursts is too short to allow the accretion of enough fresh material, necessary to trigger a Type-I X-ray burst. This suggests the presence of left-over, unburned material from the preceding burst which gets ignited in a time scale of minutes. The energies and time scales of the secondary bursts suggest a lower fraction of hydrogen compared to that estimated for the primary bursts. The persistent emission was roughly constant during the period when the Type I X-ray bursts were detected. We derive an  average accretion rate during our observations of $\dot{m}\sim\,8\,\%\,\dot{m}_{Edd}$. The spectrum of the persistent emission during these observations can be fit with a non-thermal component, indicative for the source to be in a hard state when the INTEGRAL observations were performed.}

\maketitle

\section{Introduction}
  
Type I X-ray bursts are thermonuclear runways on the surface of accreting neutron stars (hereafter NS) in Low Mass X-ray Binaries (LMXB), caused by the unstable ignition of He and/or H fuel accreted from the low-mass companion (for reviews, see, e.g., Lewin et al.\ 1993, Strohmayer \&\ Bildsten 2006).

They are short events (10--100\,s), detected as a fast rise in the source X-ray light curves, several orders of magnitude above the persistent level, followed by an exponential-like decay to pre-burst fluxes. The observed rise times are of the order of seconds, while the decays can generally last from seconds to minutes, both scales related to the relative amount of H to He burned during the  bursts (e.g., Schatz et al.\ 2001). At higher photon energies, the profiles generally show shorter exponential decays due to the cooling of the neutron star surface (see Lewin et al.\ 1993). Type I X-ray bursts show generally thermal X-ray spectra with black-body shapes (Swank et al. 1977), from a spherical region with a radius of about 10\,km and temperatures up to $\sim$3\,keV that cool during the burst decay. The energy released during these processes is typically $10^{38-39}$\,erg, and it is expected that during the flash over 90\% of the fuel accreted burns into carbon and heavier elements (e.g., Woosley et al.\ 2004). Hence, triggering a subsequent burst requires a new layer of fuel to be accumulated. The burst recurrence time can be regular or irregular on time scales of minutes to days (see Lewin et al.\ 1993). 

The burst properties depend on the composition of the accreted material, and therefore on the accretion rate. We can generally distinguish three burning regimes in a H/He accretor (see Strohmayer \& Bildsten 2006, and references therein): at low accretion rates, $\dot{m}\lesssim 900\,{\rm g\,cm}^{-2}{\rm s}^{-1}$, unstable H ignition triggers mixed H/He X-ray bursts, while at $900\lesssim\dot{m}\lesssim 2000\,{\rm g\,cm}^{-2}{\rm s}^{-1}$, H burns steadily in a shell via the CNO cycle while the pure He shell ignites. For higher accretion rates, $\dot{m}\gtrsim 2000\,{\rm g\,cm}^{-2}{\rm s}^{-1}$, H accretion is faster than H-burning and mixed H/He bursts are triggered by unstable He ignition. 

During bright Type I X-ray bursts, the peak flux can reach the local Eddington luminosity. The outward radiation pressure then equals (or exceeds) the gravitational force, pushing the outer layers of the neutron star photosphere outward. The effect observed is a decrease in the observed black-body temperature while the inferred black-body radius increases, followed by a return to the pre-expansion radius while the temperature increases. Such bursts can be used as standard candles, as they can provide a good estimate of the distance of the source (e.g., Basinska et al.\ 1984; Kuulkers et al.\ 2003).

There are some bursts showing recurrence times of the order of minutes: the so-called `short waiting-time' (SWT) bursts. SWT bursts were first reported by Lewin et al.\ (1976), who, using SAS 3 data on MXB 1743--28, detected a sequence of 3 successive bursts with recurrence times of 17 and 4 minutes, respectively. The time elapsed between successive bursts is too short to accumulate enough material to trigger the typical thermonuclear runaway, implying that the available fuel is not entirely exhausted in the first burst (Fujimoto et al.\ 1987). Exploring a sample of X-ray bursts from EXO\,0748--676, Boirin et al.\ (2007) detected burst triplets with recurrence times of 12 min. They measured that the total energy released in doublets or triplets is higher than in a normal Type I X-ray burst from that source. Later, Galloway et al.\ (2008) reported the detection of SWT bursts from 9 sources in a sample of 48 bursters. Keek et al.\ (2010) analysed doublets, triplets and even quadruples from a sample of 15 sources. They suggested several explanations for this phenomenon, one of them considering the possible scenario that the fuel of a bottom layer is ignited, leaving an unburned layer on the top that could be triggered leading to a second burst.

GS 0836--429 is a transient LMXB, first detected in 1970--1971 by {\it UHURU} (Kellogg et al. 1973) and {\it OSO-7} (as MX 0836--42; Markert et al. 1977; Cominsky et al. 1978).  A subsequent outburst from this source was detected from November 1990 until February 1991 by the All Sky Monitor (ASM) onboard Ginga (Aoki et al. 1992). The  persistent source spectrum during the outburst was described by a power law with a photon index of ∼1.5. 28 Type I X-ray bursts, with typical recurrence times of $\sim\,2$ hours were detected by Ginga during the 1990--1991 outburst. A pair of bursts with short recurrence time ($\sim$\,8\,minutes) was detected (Aoki et al.\ 1992). \
There is no report of an optical counterpart to this source, due to the high interstellar absorption in the line of sight ($\rm\,A_{\rm v}\sim\,11^{\rm m}$; Cherepashchuck 2000). Hence, it has not been possible to determine the distance to the source yet (see Belloni et al.\ 1993). 

GS 0836--429 became active again in the period January--May 2003 (Rodriguez et al.\ 2003), and  reactivated later in September 2003--June 2004. The evolution of these two outbursts was monitored by the All Sky Monitor (ASM) onboard the {\it Rossi X-ray Timing Explorer (RXTE) Satellite}.  
%\sout{The source light curve as determined by the ASM is displayed in Fig.~\ref{Fig:rxte} (top).} 
17 Type I X-ray bursts were detected by RXTE/PCA during the first outburst in 2003 (Galloway et al.\ 2008). The burst activity was also derived using {\it INTEGRAL} data by Chelovekov et al.\ (2005), who reported the detection of 24 Type I X-ray bursts by the Joint European Monitor (JEM-X) onboard {\it INTEGRAL} during the second period of activity.
GS 0836--429 was serendipitously detected during {\it INTEGRAL} observations of the Vela region (UT November 27 -- December 19, 2003; Rodriguez et al.\ 2003), as well as during a few Galactic Plane Scan pointings in April 2004 (see Table \ref{table}).

We here report on the results obtained from the analysis of all {\it INTEGRAL} observations of this source during the 2003--2004 outbursts. We characterise the system persistent emission and study the properties of 61 Type I X-ray bursts we found using JEM-X (i.e., more than double the amount of bursts reported by Chelovekov et al.). We also include a detailed study of four SWT bursts from this system.\\% , providing new results about these phenomena that can  help to better understand the ignition conditions in LMXBs. }

\section{Observations}
\begin{table*}
\centering
\caption{Observation log of GS 0836--429.}
\tiny
\begin{tabular}{l l l c c c}
		\hline\hline
		Date & MJD & ObsID & JEM-X  & ISGRI & Bursts\\
		        &			 &	& Exposure (ks) & 	Exposure (ks) & \\
		\hline
	27/11/2003 -- 11/12/2003	 & 52970 -- 52984 & 0110009/0006 & 455.4 & 999 & 58\\
	%11/12/2003 	& 0199801/1801 & \\
	02/01/2004 & 53006 & 0299820/0001 & 2.2 & 4.4 & 1\\
	24/04/2004 & 53119 & 0299827/0001 & 4.4 & 6.6 & 1\\
		\hline
		\end{tabular}
		\label{table}
		\end{table*}

{\it INTEGRAL} is an ESA scientific mission (Winkler et al.\ 2003), dedicated to fine spectroscopy (E/$\Delta$E$\sim$500; SPI; Vedrenne et al.\ 2003) and fine imaging (angular resolution: 12\arcmin\ FWHM; source location accuracy: 1--3\arcmin; IBIS; Ubertini et al.\ 2003) of celestial X-ray sources in the energy range 15\,keV to 10\,MeV with simultaneous monitoring in the X-ray range (3--35\,keV, angular resolution: 3\arcmin; Joint European X-ray Monitor, JEM-X; Lund et al.\ 2003) and in the optical (V-band, 550\,nm; OMC; Mas-Hesse et al.\ 2003). All the instruments onboard {\it INTEGRAL}, except the OMC, have coded masks.

We analysed all the available data on GS 0836--429 during the 2003--2004 outbursts collected by two of the {\it INTEGRAL} instruments (see Table \ref{table}): the {\it INTEGRAL} Soft Gamma-Ray Imager (ISGRI, part of IBIS) sensitive from $\sim$15\,keV to 1\,MeV with a total effective area of about 2600\,cm$^{2}$ (Lebrun et al.\ 2003), and the JEM-X.  IBIS has a wide field of view  (FOV, $9^{\rm o}\times9^{\rm o}$ fully coded and $29^{\rm o}\times29^{\rm o}$ partially coded; full-width at zero response, FWZR). JEM-X has a circular field of view with a diameter of about $13^{\rm o}$ (FWZR). This instrument consists of two units, which do operate simultaneously. They are sensitive in the 3--35\,keV energy range and each detector has an effective area of about 500\,cm$^2$.\
GS 0836--429 was serendipitously detected in the IBIS/ISGRI field of view during the first and second outbursts of the 2003-2004 activity phase. The source was within the JEM-X field of view only during the second outburst; this difference is due to the smaller field of view of JEM-X compared to IBIS/ISGRI. The total exposure on GS 0836--429 was $\sim$460\,ks with JEM-X and $\sim$1\,Msec with IBIS/ISGRI (see Table\,\ref{table}). 

We also used the publicly available {\it RXTE}/ASM light curves of the source (Levine, Bradt \& Cui 1996), to characterize the overall outburst evolution in the 2.5--12\,keV range. The ASM consists of three scanning wide-angle shadow cameras (SSC). Each camera has a FOV of $12^{\rm o}\times\,110^{\rm o}$ (FWZR) with a collecting area of 90  cm$^{2}$ and spatial resolution of $3'\times15'$. Counts are recorded in 1/8 s time bins for three energy bands, 1.5--3, 3--5, 5--12\,keV. 

\begin{figure*}
\centering
\includegraphics[width=0.9\textwidth]{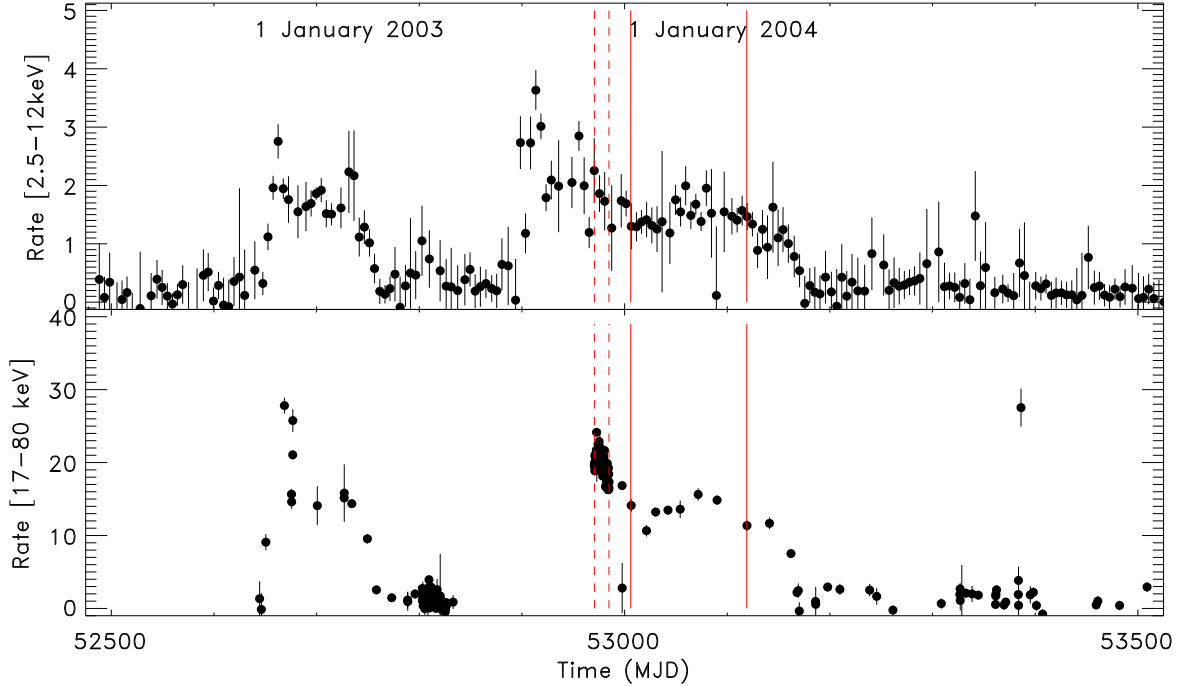}
\caption{{\it Top:} 5-day average light curve in counts per second of {\it RXTE}/ASM data covering the period of the two outbursts (January 2003 to June 2004) in the energy range 2.5--12\,keV. {\it Bottom:} 4-hrs average light curve of the {\it INTEGRAL}/IBIS/ISGRI data in the energy range of 17--80\,keV in the same period. The time between the dashed lines corresponds to the period when the source was within the JEM-X field of view, and burst searching analysis was conducted. A zoom on this time interval is provided in Figure 2, where the times of detection of X-ray bursts are shown. The solid lines indicate the times of additional Galactic Plane Scan pointings, when GS~0836--429 was again  within the JEM-X field of view. The outlier observed in the IBIS/ISGRI light curve around MJD~53400 is an instrumental artifact.  }
\label{Fig:rxte}
\end{figure*}

\begin{figure*}
\centering
\includegraphics[width=0.9\textwidth]{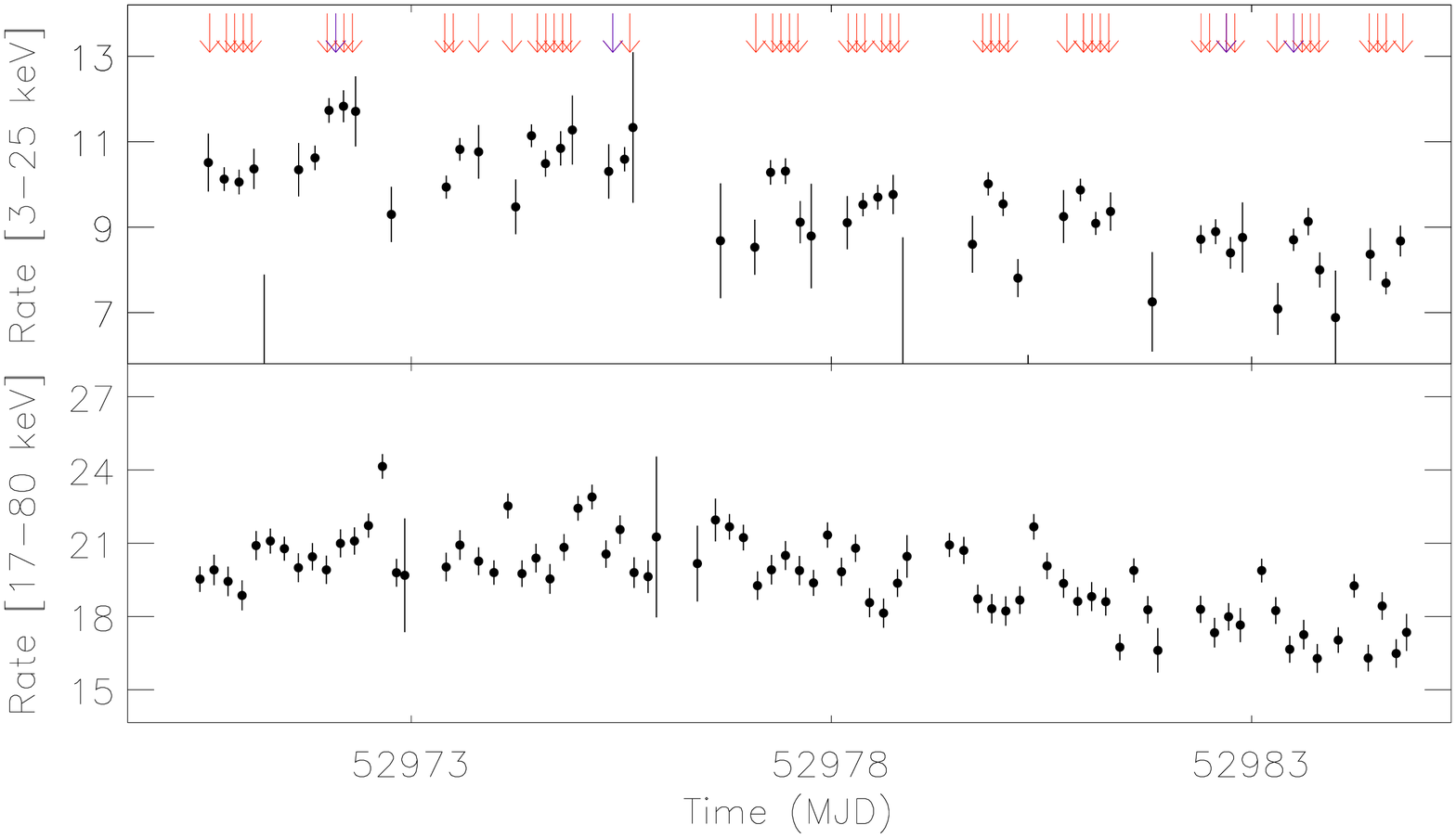}
\caption{4-hrs average light curves in counts per second of the persistent emission of GS 0836--429 built from data collected by JEM-X (3.5--25\,keV, upper panel) and IBIS/ISGRI (17--80\,keV, lower panel). The red arrows and the blue arrows shown in the upper panel indicate the time when the normal Type I X-ray bursts and the SWT bursts, respectively occurred (57, see Table~1).}
\label{Fig: four}
\end{figure*}

\section{Data reduction and analysis}
\label{sect:reduction}

The {\it INTEGRAL} data were reduced with the {\it Off-line Scientific Analysis} software (OSA) distributed by the {\it INTEGRAL} Science Data Center (ISDC; Courvoisier et al. 2003) version 10.0 released on October 18, 2012, using the OSA default parameters. The routines employed to analyse JEM-X and ISGRI data are described in Westergaard et al.\ (2003) and Goldwurm et al.\ (2003), respectively. In the next subsections we detail the analysis of the persistent emission of the source, and the detection and characterization of Type I X-ray bursts. 

\subsection{Persistent emission analysis}

The persistent emission of the source must be separated from the burst spectrum when we fit it (Swank et al.\ 1977, Kuulkers et al.\ 2002; but see, e.g., Worpel et al.\ 2013). In order to extract the spectrum of the persistent emission, we grouped the available JEM-X pointings between two consecutive bursts to build an integrated X-ray spectrum. In the cases when there was no data available between two consecutive bursts, we extracted the spectrum of the persistent emission in the pointing where the burst had been detected, using a so-called `Good Time Interval' (GTI) to exclude the burst. Due to the weakness of the persistent emission of GS 0836--429 ($\rm F_{\rm x}[5.0-200 \rm keV]\sim\,1.7\times\,10^{-9}\rm erg\,\rm s^{-1}\,\rm cm^{-2}$), we extracted the JEM-X persistent emission spectra from the mosaic image, a procedure recommended for weak sources (see Chernyakova et al.\ 2012). We processed the JEM-X data from the correction step {\sl COR} to the image creation {\sl IMA2} level. 
We then used the tool {\it mosaic\_spec} on the mosaiced image to derive the source flux in 16 energy bins, in the energy range 3--35\,keV, and generated appropriate detector and ancillary response files. The IBIS/ISGRI data were processed from the correction step {\sl COR} up to the spectrum creation level {\sl SPE}. The IBIS/ISGRI X-ray spectrum was extracted in the 15--200\,keV range using default parameters. In the analysis we have only considered those pointings for which the source was less than 4$^{\rm o}$ (JEM-X) and 12$^{\rm o}$ (ISGRI) off-axis.

To describe the persistent emission of the source we fitted the JEM-X and IBIS/ISGRI spectra simultaneously with a model consisting of a power-law with a cut-off at high energies ({\sl{cutoffpl}} in \textsf{XSPEC} v.12; Arnaud 1996)  modified by photo-electric absorption ({\sl phabs}). We fixed the column density to $N_{H} = 2.7\times10^{21}\,{\rm cm}^{-2}$ (Chelovekov et al.\ 2005). In order to account for the relative JEM-X/ISGRI calibration uncertainties we introduced an additional multiplicative factor in the  fits. Due to calibration uncertainties below 5 keV, we neglected the JEM-X energy bins below that value to fit the spectra, so that the energy range covers from 5 up to 200 keV. The luminosities derived from this analysis are provided in Table A.1. Note that for those cases when the persistent emission spectra was extracted using $\leq\,2$ pointings it was not always possible to obtain a satisfactory spectral fit to the system persistent emission.

The total average source spectrum for the dataset analyzed in this work was also extracted and fitted.
To estimate the absolute persistent fluxes before a burst, we first extracted the flux in the 5--35 keV band using the JEM-X and ISGRI persistent spectra (see analysis procedures above) and extrapolated that to 0.1--200 keV range by multiplying by a bolometric factor. This factor was determined by using the total average source spectrum. We used the persistent emission fluxes to derive the burst parameter $\alpha$, used to characterize the composition of the fuel triggering the thermonuclear runaway (see Section 4.2).

\subsection{X-ray bursts analysis}
\subsubsection{X-ray bursts light curve and burst detection}

We applied X-ray burst searching routines to the 5-s binned source light curves extracted in the 3.5--25 keV energy band. The potential onset of a burst was flagged when, in a time bin, the difference between the source count rate and the average source count rate in that pointing exceeded four times the light curve noise (measured as the standard deviation of the pointing source count rate). The subsequent burst decay was fit with an exponential function ($\exp{(-t/\tau)}$). From the fits to the burst light curve we derived the basic burst parameters: burst peak count rate, burst duration, burst e-folding decay time $\tau$, and burst recurrence time. Burst starting and end times were also identified: they are defined as, respectively, the time when the intensity increases above 10\% of the burst peak count rate, and the time when the intensity decreases again below 10\% of it. The start and end times were used to generate GTI files which allowed the creation of images and spectra covering the burst duration. The images were checked to verify that the event indeed originated from GS~0836--429, and not from the other transient source GS~0834--430, at an angular distance of 24' from our target. GS~0834--430 is a high mass X-ray binary (HMXB) with an accreting pulsar (see Miyasaka et al. 2013).

We also extracted burst light curves with a time step of 2\,s to built an average burst profile to estimate the burst average rise time, peak count rate and exponential decay time.

\subsubsection{Burst spectral analysis}

Applying the above described GTIs, we processed the JEM-X data from {\sl COR} to {\it SPE} steps using 16 energy bins in the energy range 3.5--35\,keV. The spectra of the  weakest bursts in our sample, i.e., those showing shorter durations ($<$40\,s) and/or lower peak intensities ($<$65 counts\,s$^{-1}$), were extracted using 8 energy bins in the same energy range. We also generated the appropriate detector and ancillary response files. Due to uncertainties in the flux reconstruction at high off-axis distances, we did not extract spectra for those bursts detected at $\geq$4$^{\rm o}$ off-axis (see Table A.1).  Due to the uncertainties in the calibration below 5 keV, and bad response from the detector above 25 keV, we fit the burst spectra from 5 -- 25\, keV. 

The spectrum of the persistent emission in the same energy range was subtracted from the total burst emission, assuming that the persistent emission during the burst remains unchanged. Although Worpel et al.\ (2013) suggested  that the persistent emission may increase during bursts, the statistics of the JEM-X spectra, are not sensitive enough to include this effect. 
Our net-burst spectra are well described by a black-body component modified by the interstellar absorption, which is dominated by photo-electric absorption. Its effect appears as a low-energy cut-off. 
We fixed the photo-electron absorption at $N_{H} = 2.7\times10^{21}\,{\rm cm}^{-2}$. To get an estimate of the neutron star photospheric radius, we used the \textsf{XSPEC} model {\sl bbodyrad}. Note, however, that this analysis may contain systematic errors, more important in the tail of the burst where the fluxes are low. As a consequence the inferred radii obtained may become artificially small (see Lewin et al.\ 1993). Also, in the fitting process the radius is derived  from the burst colour temperature, and not from the effective temperature, so that the values are underestimated (e.g., van Paradijs 1982; London et al.\ 1984).
From these spectral fits we measure the burst fluxes and derive the burst fluences.

\section{Results}

We report in this Section the results of the analysis of the persistent emission during the 2003--2004 outbursts (Sect.~4.1), and provide the X-ray burst parameters obtained in the analysis of our sample of 61 X-ray bursts (Sect.~4.2, see Table \ref{table:longtables}). We note that 24 of these bursts were previously reported by Chelovekov et al.\ (2005).\footnote{The report by Chelovekov et al.\ (2005) is not detailed enough to assess why their sample of bursts is smaller than ours.} An overview of the properties of the single bursts and the persistent emission around the burst is also given in Sect.~\ref{sect:persistent}. We searched for possible correlations between the parameters of the bursts and the source persistent emission, and derived the distance and an estimate of the mass accretion rate. Finally, we describe in detail the properties of the SWT bursts detected in this work in Sect.~4.3. 

\subsection{Persistent emission}
\label{sect:persistent}

The {\it RXTE}/ASM 5-day average light curve of GS 0836--429 (2.5--12\,keV) during the the 2003--2004 outbursts is presented in Fig.~\ref{Fig:rxte} (top). Contemporaneous IBIS/ISGRI measurements in the 17--80 keV energy band are also presented in this figure (bottom). Both outbursts display a similar morphology: a big flare in soft and hard X-rays followed by a plateau lasting several weeks until the source decays to quiescence. The outburst durations are, however, different: the first outburst lasted about 100 days in soft X-rays, while the second one had a duration of about 250 days in the same energy range (see Fig.~\ref{Fig:rxte}). The onset of the outbursts are different as well, i.e., the onset of the first outburst is brighter in hard X-rays, while the onset of the second outburst is brighter in soft X-rays. The plateaus of the two outbursts have comparable fluxes. The observed light curves are indicative that the second outburst started earlier in the soft X-rays, than in hard X-rays, as observed in other LMXB transients (e.g., Lewin et al. 1993). %The outlier observed in the IBIS/ISGRI light curve is an instrumental artifact. 
A closer  view of the epoch when the source was within the JEM-X FOV and Type I X-ray bursts were detected is shown in Fig.~\ref{Fig: four}. In this figure we show the system light curve in two energy bands (3--25\,keV and 17--80\,keV); the rates in both energy bands decrease during our observations by about 20\%. 

\begin{figure}[h!]
\centering
\includegraphics[width=6cm, angle=-90]{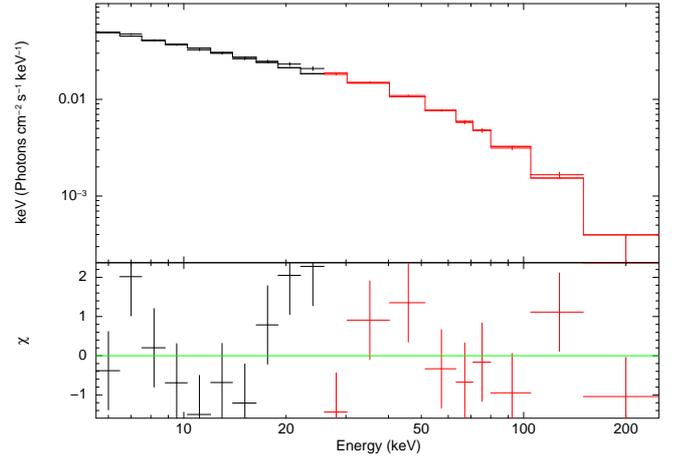}
\caption{Top: Overall, average persistent emission spectrum as seen by JEM-X (5--25\,keV) and IBIS/ISGRI (25--200\,keV). The spectrum has been fitted with a cut-off power-law with photon index of 1.5 and a high energy cut-off at 57\,keV (see text). Bottom: Residuals from the above fit in units of $\chi$.}
\label{Fig:spectra2}
\end{figure}

The overall, average source persistent emission was detected up to about 100\,keV with IBIS/ISGRI (see Fig.~\ref{Fig:spectra2}). Its spectrum was fitted by a power-law model with a cut-off at high energies. The photon index ($\Gamma$) derived from this fit (with a $\chi_{\rm red}^{2}$ of 1.7 for 22 degrees of freedom, d.o.f.) is 1.52$\pm$0.05, with a cut-off energy of 57$\pm$4\,keV. The best estimate of the bolometric correction was determined from a joint analysis of the average JEM-X and ISGRI spectra (see Sect.~3.1): we obtained a value of $1.16\pm0.16$. 

We also performed spectral fits to the absorbed persistent spectra (3--200 keV) around the burst detections. The values for the fluxes derived in these fits are typically between $1.4\,\times10^{-9}$ and $2.4\,\times10^{-9}{\rm\,erg\,s^{-1}cm}^{-2}$. The flux decreased by a factor of about 1.3 in 15 days (see Table~\ref{table:longtables}). 
%1.2 to $1.8\,10^{-9}\erg\cm^{-2}s^{-1}$ 

\begin{figure}[h!]
\centering
\includegraphics[width=0.5\textwidth]{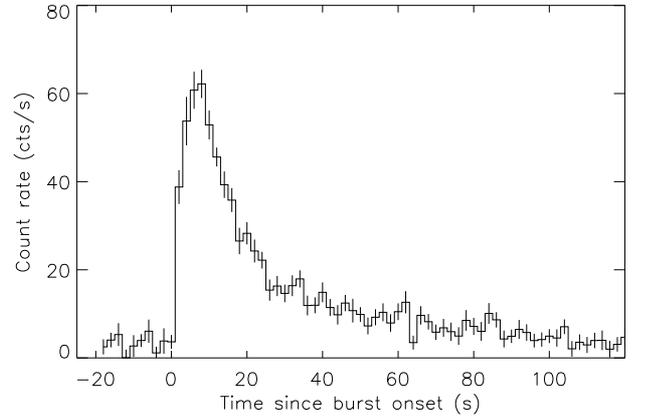}
\caption{Average burst profile (3.5--25\,keV) of GS 0836--429 built from all standard X-ray bursts in our sample (i.e., excluding secondary bursts in SWT events) detected $\leq$4$^{\rm o}$ off-axis. Time resolution is 2\,s. The average burst profile shows a fast rise ($\sim$\,7\,sec) followed by a long exponential tail (e-folding decay time $\sim$\,29\,s) to pre-burst levels.}
\label{Fig:average}
\end{figure}
 
\subsection{Type I X-ray bursts}

\subsubsection{Burst light curves}

\begin{figure}[h!]
%\centering
\includegraphics[width=0.5\textwidth]{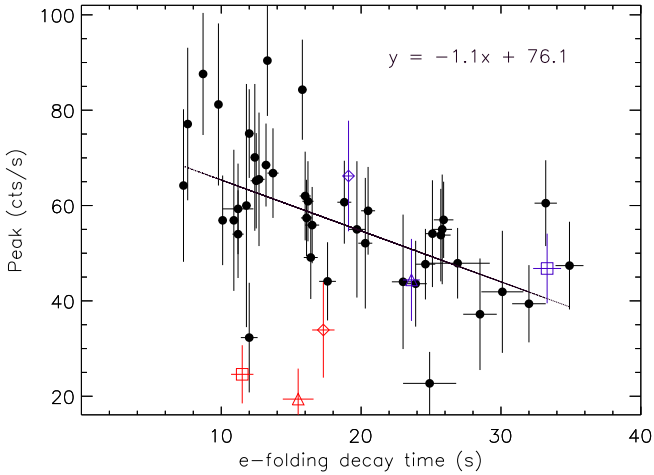} 
\caption{Relation between burst peak count rate and  e-folding exponential decay time for the bursts in our sample. To avoid systematic effects related to flux reconstruction at high off-axis distances, only the parameters of the bursts detected at off-axis distances $\le\,4^{\rm o}$ are shown. Blue symbols correspond to primary bursts in SWT doublets. Red symbols belong to secondary bursts in SWT doublets. Each pair of bursts in a SWT pair are represented using the same symbol shape. For standard bursts in the sample, the peak count rate  decreases as the e-folding exponential decay increases.  The solid line indicates the linear fit with a $\chi_{\rm red}^{2}$ of 1.4 for 40 d.o.f., excluding bursts in SWT doublets.}
\label{Fig:corr}
\end{figure}

Using JEM-X we detected 61 Type I X-ray bursts during the 2003--2004 outburst (see Table A.1). The X-ray burst light curves show a fast rise, which last on average $7\pm\,2$\,s, followed by a long exponential decay (see Fig.~\ref{Fig:average}). The average burst duration defined as the time elapsed since the burst count rate exceeds 10\% of the peak count rate until it decays to  10\% of the burst peak count rate is $49\pm\,2$\,s.  The average e-folding exponential decay time is $19\pm\,1$\,s. However, we found bursts with exponential decay times in the range $\simeq$7 to $\simeq$47\,s. The average of the net-burst peak count rate is $55\pm12$\,cts\,s$^{-1}$, with burst peak count rates ranging  from 22 to 101\,cts\,s$^{-1}$ (3.5--25\,keV). 
We find that bursts with higher peak count rates tend to display faster decays (see Fig.~\ref{Fig:corr}; excluding SWT doublets). In Fig.~\ref{Fig:corr} we also display the burst peak count rates and e-folding decay times for bursts in SWT doublets. Primary bursts in SWT doublets occupy the same regions in this plot as the typical single bursts, while secondary bursts in SWT doublets tend to show lower peak count rates and faster decays. Since for the weakest bursts it is more difficult to fit the exponential decay and this can bias our results we have tested whether this relation is real. First, we explored the effect of adding extra noise to the burst profile and we confirmed that it does not affect the exponential decay fit. Secondly, we tested if a decrease in burst strength due to the burst being at higher off-axis angles affects the exponential decay fit, but no correlation has been found indicating that the relation is real. We have tested whether a linear fit describes better the data than a constant fit performing an F-test, and we can reject the constant model with 0.99 probability.

\begin{figure}[h!]
\centering
\includegraphics[width=0.5\textwidth]{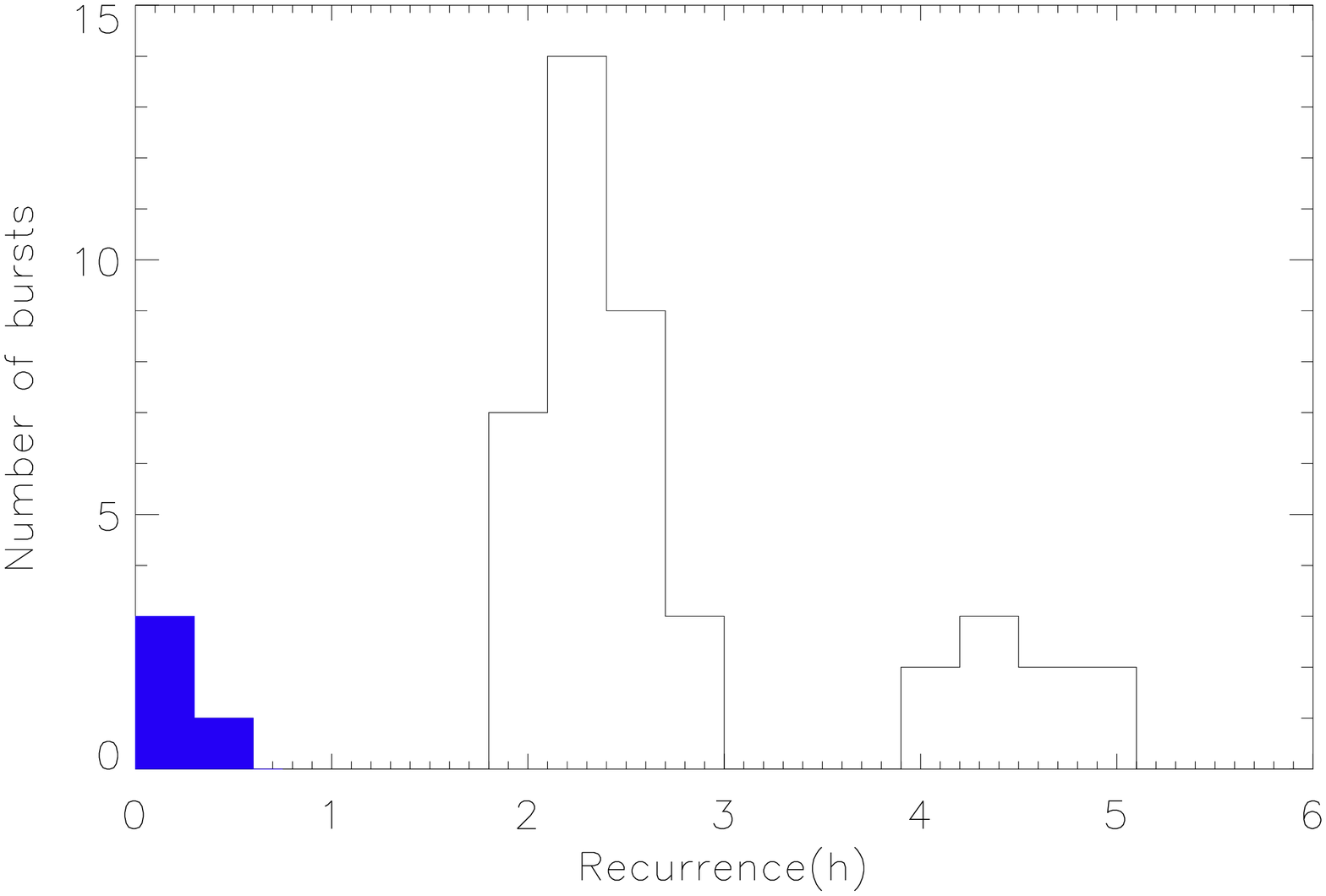} 
\includegraphics[width=0.5\textwidth]{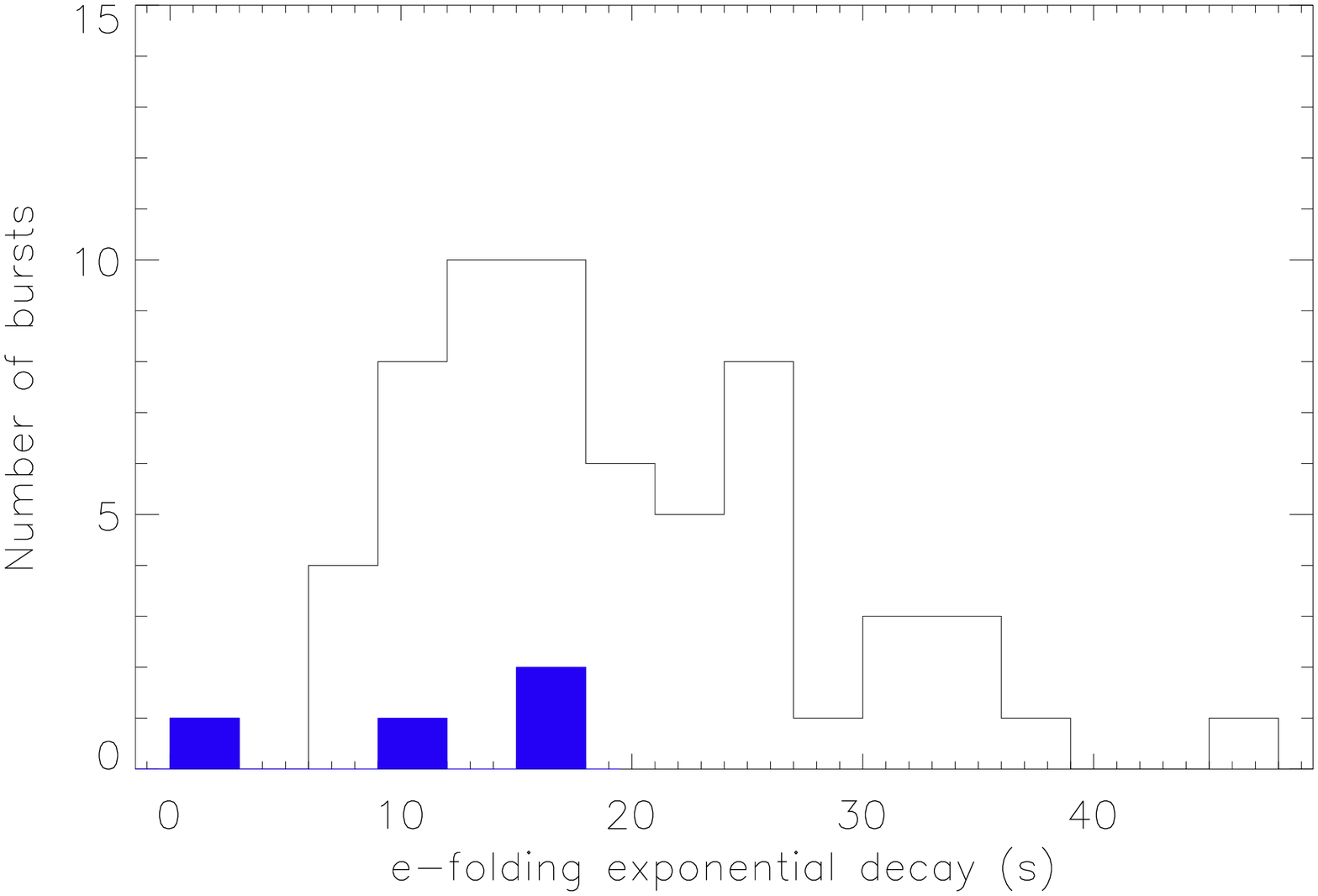}
\caption{{\it Top}: Distribution of burst recurrence times below 6\,hr. The black line shows the distribution of single bursts in our sample, with a peak value around 2.3\,hr. Secondary bursts in SWT doublets are represented by the solid blue histograms. Recurrence times around 4\,hr and above are due to gaps in the observations.  {\it Bottom}: Distribution of e-folding exponential decay times. The black line shows the distribution of single bursts in our sample. The data suggest a maximum in the distribution at around 15\,s. Secondary bursts in SWT doublets are represented by the solid blue histogram.}
\label{Fig:histograms}
\end{figure}

We find a quasi-periodic waiting time between two successive bursts. The average value of the recurrence time is $2.3\pm\,0.5$\,hr. This value is derived excluding the secondary bursts in SWT doublets, but takes into account the data gaps between two bursts events caused by {\it INTEGRAL}'s 5x5 dithering observing pattern: 1 source on-axis pointing, 24 off-source pointings, with a 2.17 degrees step. In the calculation of the average recurrence time, we have divided by two the burst recurrence times in the range 4--6\,hr (see Fig.~\ref{Fig:histograms}).

In our data we do not find a significant correlation between the burst peak count rate and burst recurrence time. One would expect that for longer waiting times between two bursts, the neutron star would have accreted more material, and, therefore, the bursts would be brighter and more energetic (if the triggering conditions are the same; see Lewin et al.\ 1993).  We do not find any correlation, either, between persistent emission and burst recurrence time.

\subsubsection{Fits to the net-burst spectra}

Fits to the net-burst integrated spectra allowed to derive the average burst apparent temperature and fluence. We also built an average, overall burst spectrum from those bursts detected at an off-axis distance less or equal than four degrees to derive more refined parameters. Secondary bursts in SWT doublets were excluded from this selection. 

We fitted the burst spectra with a black-body model subjected to a fixed interstellar absorption of $\rm N_{\rm H} = 2.7\times10^{21}\,{\rm cm}^{-2}$. 
We found apparent black-body temperatures, kT, in the range of 1.2--2.4\,keV, with an average value $2.2 \pm 0.4$\,keV. Fluences were determined using the time integrated flux of each burst in the energy range 3.5--25 keV, and were in the range $0.9-5.8\,\times\,10^{-7}\rm\,erg\,cm^{-2}$, with an average of ($2.3\,\pm 0.7\,)\times\,10^{-7}\rm\,erg\,cm^{-2}$.
The fits showed $\chi_{\rm red}^{2}$ values in the range 0.1--1.3 (for 12 d.o.f.). The very low $\chi_{\rm red}^{2}$ values can be explained by several reasons: 1) weakness of the burst; 2) burst detected at high off-axis values, which leads to uncertainties in the flux reconstruction; 3) uncertainties in the determination of the persistent emission (used in the fits to the net-burst spectra), caused by a limited amount of data between two consecutive bursts.

We explored possible correlations between burst fluence, burst peak count rate and burst recurrence time. We do not find significant correlations between these parameters. 

We used one of the brightest bursts in our sample detected at an off-axis distance of $0.8^{\rm o}$ (i.e., the one occurring on MJD 52977.46563) to estimate a representative apparent photospheric radius of the neutron star. We found a value of $9\pm\,3$\,km, consistent with a canonical neutron star radius of 10 km. For the rest of the bursts we found an average inferred black-body radius of $4\pm2$ km (but see Sect.\ 3.2.2).

Using the bolometric persistent flux, burst recurrence time and burst fluence in the energy range of 3.5 up to 25 keV, we determined the parameter $\alpha$. This dimensionless number is used to compare the amount of gravitational energy released  by the accretion of the fuel that powers an X-ray burst with the nuclear energy released during the X-ray burst burning of that fuel. We only determined this coefficient for bursts which allowed unambiguous recurrence time determinations (i.e., those with  recurrence times of $\simeq$2\,hr, see below). We derive $\alpha = 49\pm$3, indicative of  a regime of mixed H/He fuel (see, e.g., Galloway et al.\ 2008). 
We explored correlations between $\alpha$, persistent flux, burst recurrence time and burst fluence. No significant correlations were found. However, $\alpha$ is seen to decrease when the fluence increases; this is as expected since they are inversely proportional.

Although we did not find any evidence for Eddington-limited radius-expansion bursts, we derive an upper limit of the distance to the source using the brightest X-ray burst of our sample (i.e., the one on MJD 52983.82, see Table A.1), assuming the peak is below the Eddington limit. We used the burst peak count rate and a conversion factor between count rate and flux in the energy range of 3--25\,keV, i.e., 148 cts$\rm s^{-1}$ equals $2.9\times\,10^{-8}\,\rm erg\,s^{-1}\rm cm^{-2}$ (J.~Chenevez, 2015, private communication). We derive a maximum peak flux for that burst of $\rm F_{\rm x}[3-25\,\rm keV]=1.978\times\,10^{-8}\,\rm erg\,s^{-1}\rm cm^{-2}$.\footnote{This is the brightest burst observed in our sample; the brightest burst which occurred in the previous outburst and detected by {\it RXTE}/PCA had a lower flux of $F_{\rm x}[3-20\,\rm keV]=1.63\times\,10^{-8}\,\rm erg\,s^{-1}\rm cm^{-2}$ (e.g., Chelovekov et al.\ 2005).} Assuming canonical neutron star parameters (mass of 1.4\,M$_{\odot}$ and radius of 10\,km) and hydrogen-rich accreted material, we then find an upper limit on the distance to the source of about 9.2\,kpc.

Using the upper limit to the distance, we estimated the local mass accretion rate of $\dot{m}\lesssim\,8\,\%\,\dot{m}_{\rm Edd}$, again assuming a canonical mass and radius for the neutron star (see Eq.~2 from Galloway et al.\ 2008). 

\subsection{Short Waiting Time (SWT) bursts}

We detected four groups of burst doublets with recurrence times of 9--20 minutes within the doublets (see Fig.~7). The parameters of these bursts are provided in Table \ref{table2}. For each of these bursts we also determined the $\alpha$ parameter. The fluence was derived using fits to the net-burst X-ray spectrum, whenever possible:
if the burst was too weak to allow the extraction of the integrated burst spectrum, we estimated the fluence by converting the burst peak count rate to flux (see Sect.~4.2) and multiplying it by the e-folding exponential decay time.

Primary bursts in SWT doublets display light curves similar to those of single bursts in our sample, but secondary burst are $\sim 60\,\%$ weaker with fluences $0.6-0.9\,\times\,10^{-7}\rm erg\,\rm s^{-1}\rm cm^{-2}$ and display durations from 11.5 up to 15.5 seconds, shorter than single bursts in our sample. The average burst profile of these four SWT bursts has a peak count rate of $22\pm\,2\, \rm counts\,\rm s^{-1}$ and an e-folding exponential decay $5.4\pm\, 0.8\, \rm s$. We do not find substantial differences on the waiting times before and after SWT events, compared with the single bursts in our study. Primary bursts in SWT doublets tend to show $\alpha$ values from 70.4 up to 127, larger than the typical $\alpha$ values found for this work that are $\sim 49$, which suggest that the accreted material is not entirely burned in the burst (Galloway et al.\ 2008). The secondary bursts in the SWT doublets, on the other hand, display $\alpha$ values $\sim 20$, smaller than the average $\alpha$ from our sample, suggesting that fuel left over from the primary burst is contributing to the burst fluence of the secondary burst.

\begin{table*}
\centering
\tiny
	\begin{tabular}{l c c c c c c}
	\hline\hline
	SWT $\#$ & MJD & time to & Peak & $\tau$ (s) & E$_{\rm b}$ & $\alpha$ \\
	 & & previous burst (h) & ($\rm cts\,\rm s^{-1}$) & & ($\times\,10^{-7}\rm erg\,\rm s^{-1}\rm cm^{-2}$) & \\
	\hline
	I & 52972.05397 & 2.38 & 47 & 33.3 & 2.4 & 70.4\\
	 & 52972.06034 & 0.15 & 25 & 11.5 & 0.6 & 18.9\\
	 & & & & & & \\
	II & 52975.37217 & 10.53 & 66 & 19.1 & 1.8 & 90.7$^{*}$\\
	 & 52975.38611 & 0.33 & 34 & 17.3 & 0.9 & 26.0\\
	 & & & & & & \\
	III & 52982.65130 & 2.87 & 37 & 18.2 & 1.6 & 127.3\\
	 & 52982.65917 & 0.19 & 23 & -- & -- & --\\
	 & & & & & & \\
	IV & 52983.50750 & 4.49 & 44 & 23.6 & 1.9 & 85.9$^{*}$\\
	 & 52983.51408 & 0.16 & 19.4 & 15.5 & 0.6 & 19.3\\
		\hline
	\end{tabular}
	\caption{Parameters for our double SWT bursts.The $^{*}$ in the last column indicates that these values have been estimated assuming the regular recurrence found in this work of 2.3 h, since the long times to the previous burst as displayed in the 2nd column is due to data gaps in the observations.}
	\label{table2}
\end{table*}

\begin{figure*}
\centering
		\subfigure[]{%
        \includegraphics[width=7cm]{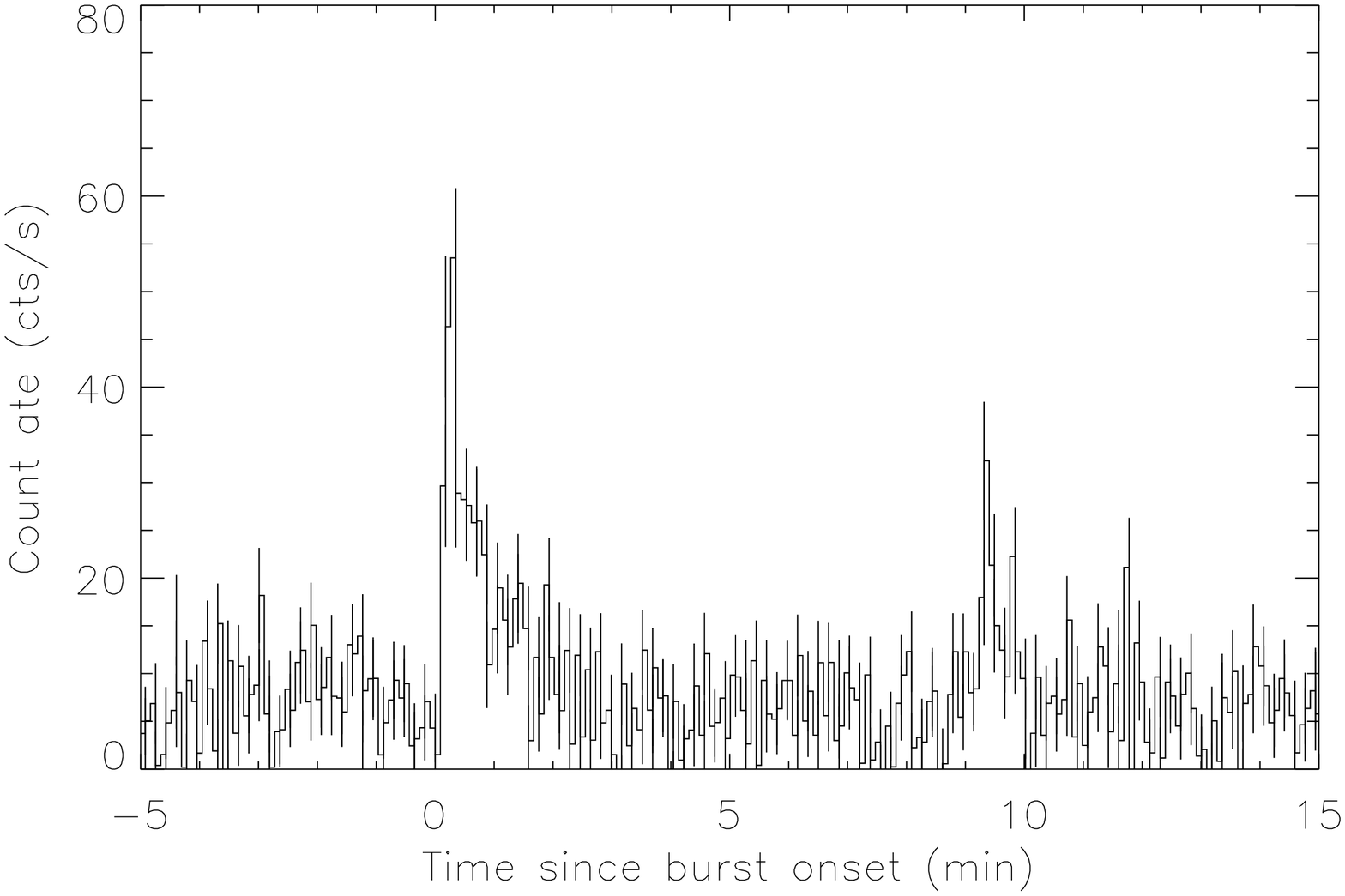}
        }
		\subfigure[]{%
		\includegraphics[width=7cm]{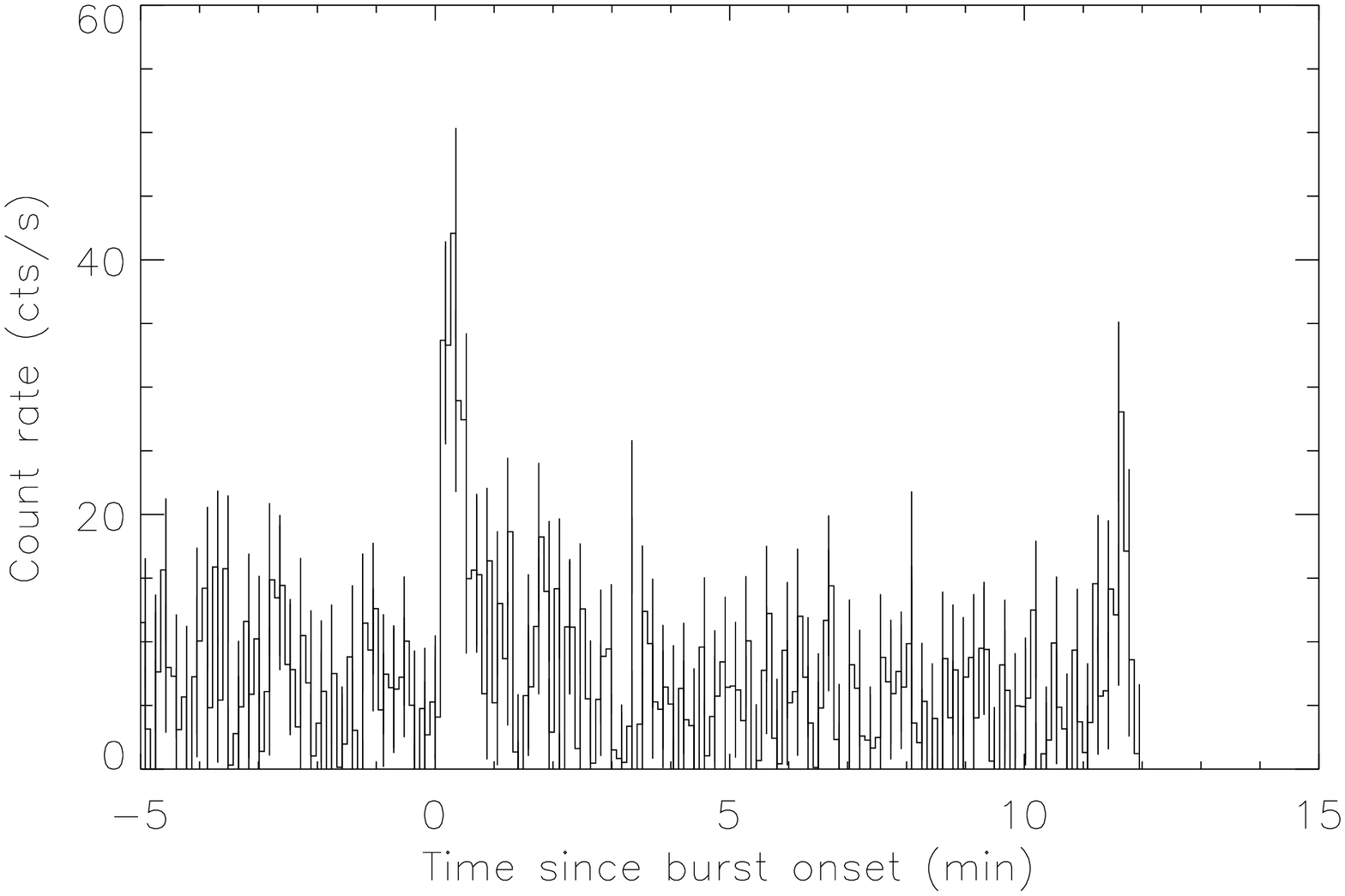}
		}\\
		\subfigure[]{%			
		\includegraphics[width=7cm]{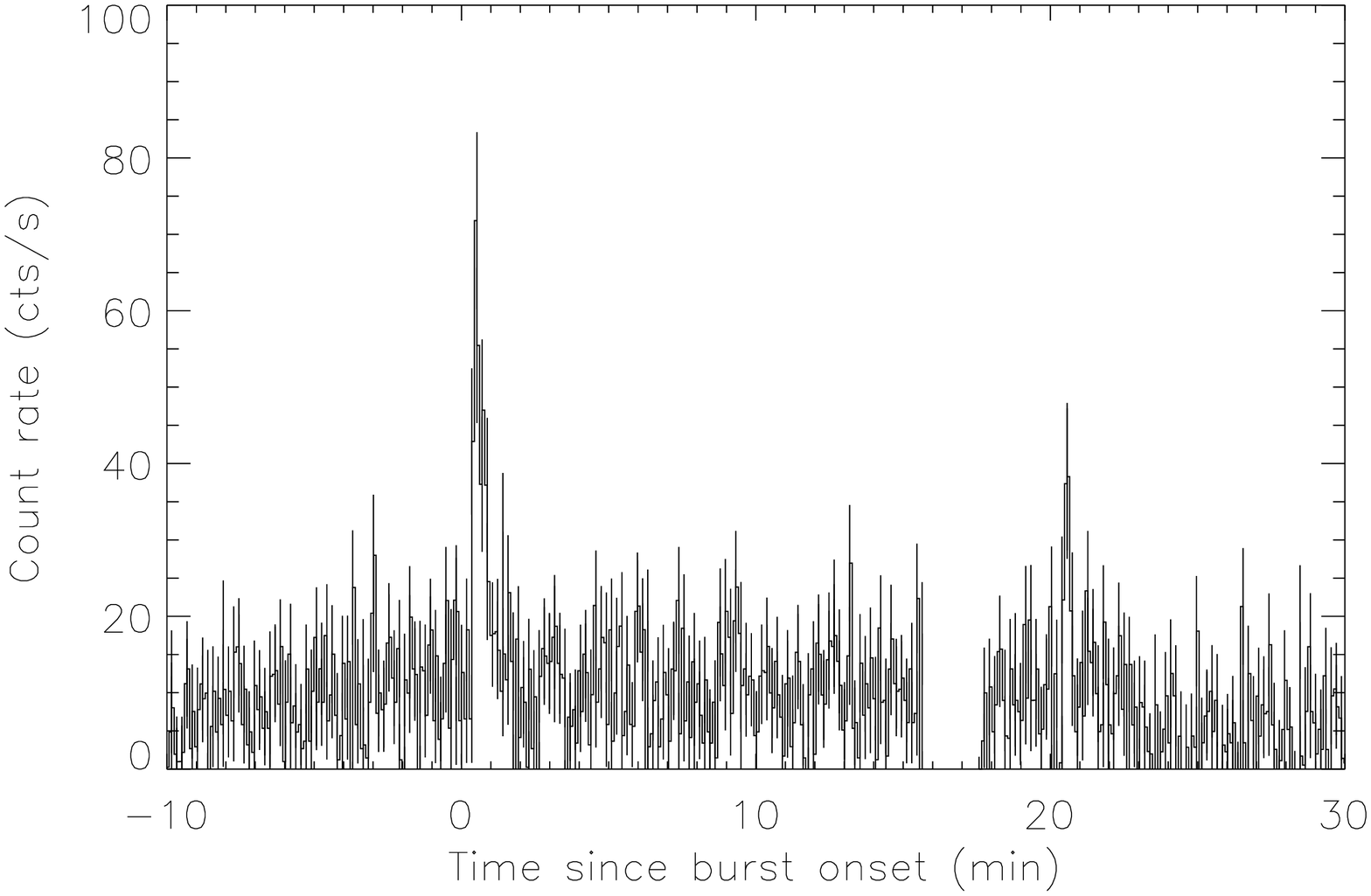}
		}
		\subfigure[]{%
		\includegraphics[width=7cm]{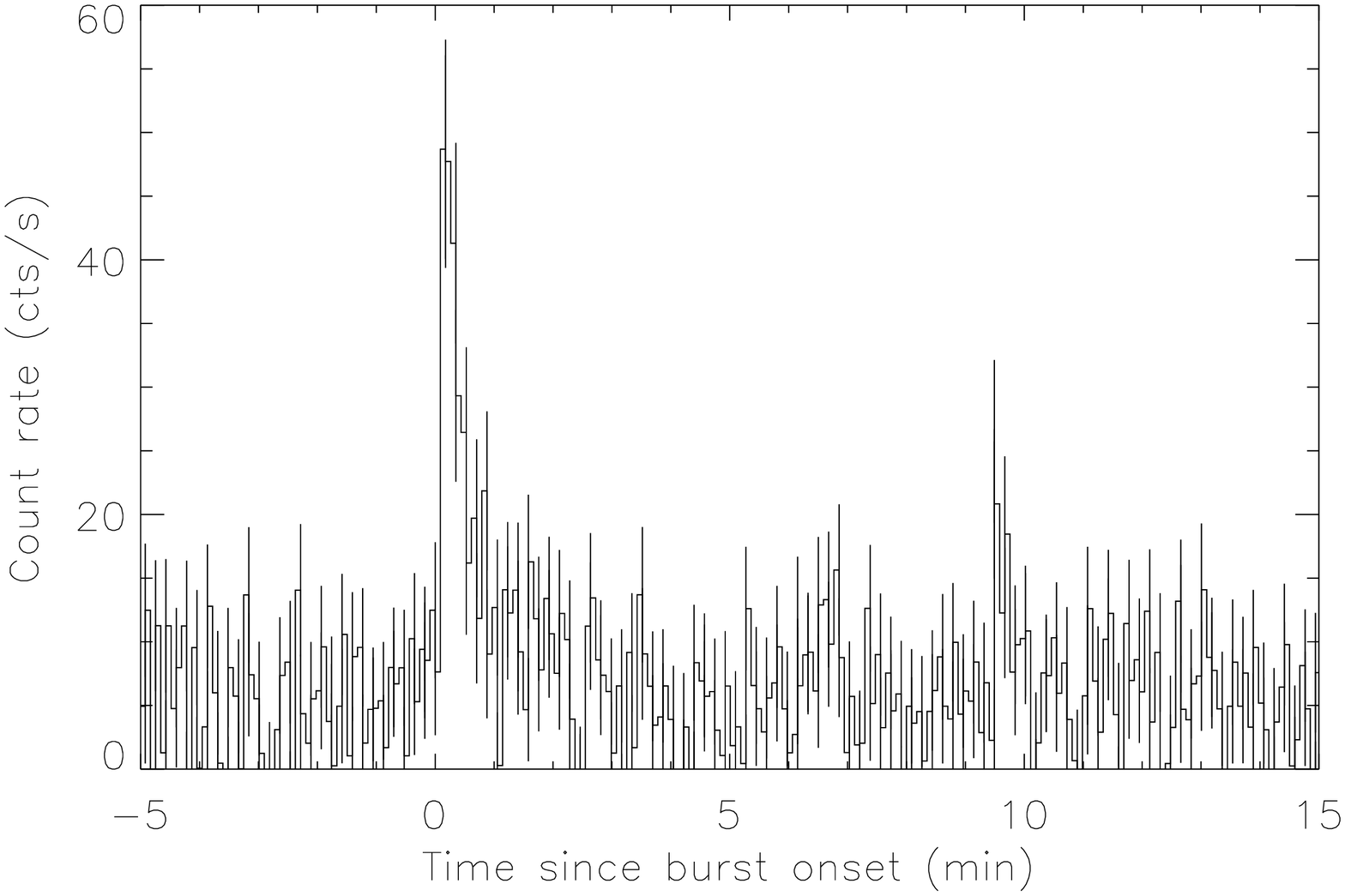}
		}\\
\caption{JEM-X light curves (3--25 keV) of the four SWT pairs studied in this work. In the four the cases the primary bursts are brighter than the secondary bursts. The secondary bursts do not display the long exponential tails detected in the standard bursts in this work, indicating a lower H content in the burning fuel. The parameters for the bursts in these pairs are provided in Table 2. {\it (a)}~SWT~I, {\it (b)}~SWT~II, {\it (c)}~SWT~III, {\it (d)}~SWT~IV.}
\label{Fig:double}
\end{figure*}

\section{Discussion}

\subsection{GS~0836--429 outburst and Type~I X-ray burst properties}

Using the available data, we have studied the properties of the transient low-mass X-ray binary system GS~0836--429 over the period 2003--2004, when two outbursts were detected in X-rays. We have analyzed in more detail the period November 27 to December 19, 2003, when the source was serendipitously detected by the JEM-X and IBIS/ISGRI instruments onboard {\sl INTEGRAL}. The average system bolometric luminosity during the period of the outburst where the source was in the field-of-view of the JEM-X instrument was $1.7\times\,10^{37}\,\rm erg\rm s^{-1}$ (assuming a distance of 9.2\,kpc). The luminosity was observed to decrease by $\sim$ 20\%\ 
during this period. The average persistent spectrum of the system was fitted with an absorbed power-law ($\Gamma\,\sim 1.5$) modified by a cut-off at high energies (at $\sim$\,57\,keV).
The non-detection of a soft component suggests that the source was in a hard state during that part of our observations.

Using the available JEM-X data we detected 61 Type-I X-ray bursts. We fitted the burst light curve profiles  %and burst spectra 
and derived the basic burst light curve parameters: rise time (on average $\simeq$7\,s), peak luminosity of $1.1\times\,10^{38}\,\rm erg\,s^{-1}$ at 9.2\ kpc,
and e-folding exponential decay time (on average $\simeq$19\,s).
Fits to the burst spectra allowed to derive the apparent burst temperatures ($\simeq$2\,keV), as well as the total energy released during the bursts (i.e., fluence $\simeq7\times\,10^{37}\rm erg\,\rm s^{-1}$).

Similar burst durations, burst fluences and value for the $\alpha$ parameter have been reported by Galloway et al.\ (2008), using a sample of 17 PCA X-ray bursts that occurred during the first 2003 outburst. Chelovekov et al.\ (2005) reported from their sample, which overlaps with ours, also similar burst profiles and fluences. They, however, estimated an $\alpha$ value of 144, suggesting that the X-ray bursts were triggered by pure He fuel (but they concluded that the shape of the bursts was consistent instead with H/He mixed bursts). We note that they used an average burst duration of 12.3\,s, which is shorter than the average burst duration of $\simeq\,50$ s derived in this work, and explains the higher value of $\alpha$. Comparing our results with those of Galloway et al.\ (2008), we conclude that the burst activity in terms of time scales and energetics is similar in both outbursts and, therefore, the burst triggering mechanism and fuel composition is the same.

We derive an upper limit to the distance to the source of about 9.2\,kpc, consistent with previous distance determinations by  Chelovekov et al.\ (2005) and Galloway et al.\ (2008). We derive an apparent neutron star radius of $\sim$\,9\,km, consistent with the canonical neutron star radius, which was inferred from fits to the spectrum of a bright burst in our sample. We derived the local accretion rate onto the neutron star to be about $\dot{m}\lesssim\,8\,\%\,{\rm\dot{m}_{Edd}}$. This would correspond to case 2 in Fujimoto et al. (1981), where it is predicted that for above approximately 5\,\%\,$\dot{m}_{\rm Edd}$ H is accreted faster than it can be consumed by steady burning (limited by the rate of $\beta$ decays in the CNO cycle), and He ignites unstably in an H-rich environment.
The long tails displayed by the burst light curves, and the value of $\alpha$ derived for the bursts analyzed in this study ($\simeq$49), are compatible with this assumption.

We derive a quasi-periodic waiting time of $2.3\pm\,0.5$\,hr between two successive bursts. Quasi-periodic bursting has been found in other sources, the best example being GS~1826--24, which consistently displays quasi-periodic Type-I bursts with recurrence times of $\sim\,3.56-5.74$\,hr (see, e.g., Cornelisse at al. 2003, Galloway et al. 2004, and references therein; but see Chenevez et al.\ 2015). Analyzing {\sl BeppoSAX} data of a sample of 9 galactic X-ray bursters, Cornelisse et al. 2003 found that for luminosities L$_{\rm X} \lesssim 2 \times\,10^{37}\,\rm erg\rm s^{-1}$, bursts occur quasi-periodically, and the burst rate increases linearly with accretion rate. The persistent emission of GS~0836--429 inferred during our observations, L$_{\rm X} = 1.7 \times\,10^{37}\,\rm erg\rm s^{-1}$, is consistent with this accretion regime. The non-detection of significant correlations between burst recurrence time and flux (and hence accretion rate) is probably due to the small variations in accretion rate during our observations.

\subsection{SWT bursts}

We found four pairs of SWT bursts. The measured recurrence times  in these burst doublets (9--20 minutes) are too short to allow the accretion of enough fresh material to trigger the secondary bursts. This implies that they are probably triggered by fuel left over after the primary burst. This  hypothesis is supported by the $\alpha$ values derived for the bursts in the SWT doublets in our sample (see Table \ref{table2}). 
The primary bursts in  SWT doublets  display $\alpha$ values from 70 up to 127, higher than the average $\alpha$\,$\sim$\,49 derived for the single bursts in this work. The secondary bursts, on the other hand, display lower $\alpha$ values ($\sim\,20$) than the average in our sample.  If the accreted fuel is not completely consumed in the primary burst, the observed fluence will be lower than expected, and  $\alpha$ will be in excess of the expected value (Galloway et al. 2008). The left-over fuel will be burned during the second burst, for which the observed fluence will be higher than expected, and  $\alpha$ will be smaller than for single bursts. The profiles of the secondary bursts lack the long tails detected in the single bursts in this study,  caused by {\sl rp}-process. This may indicate that the fuel burned in secondary  bursts  have a lower H content than primary bursts and standard bursts in this work. A study by Boirin et al.\ (2007) showed that in EXO 0748--676 a doublet was always more energetic than a singlet. In our case there are normal, single bursts showing higher fluences than some doublets.

Galloway et al.\ (2008) found indications that SWT groups happen when the persistent flux is between 2\,\% and 4\,\% of the Eddington limited flux, and occur predominantly in H-rich accretors. However, Keek et al. (2010) observed SWT bursts over the entire range of mass accretion rates where single bursts are observed. 
Still, SWT bursts have not been detected from ultra-compact X-ray binaries. Since the neutron star in these systems are thought to accrete He-rich material, this indicates that the H burning processes (hot CNO cycle, $\alpha$p-process, $rp$-process) may play a role in the occurrence of SWT bursts (see Keek et al. 2010). 
Studying a sample of 44 H-rich accretors, they found, however, that only 15 of them displayed SWT events. So accretion of H-rich material is a requirement for the production of SWT events, but not a sufficient condition.
Studying the position of H-accreting sources in the color-color diagram, Keek et al. (2010) found that SWT bursts are restricted to the so-called `island state', while the occurrence of single bursts can also occur in the `banana' branch. Note that, although the regular burster GS 1826--24 spends most of its time in the `island' state (see Chenevez et al.\ 2015), there are no reports of SWT bursts from this source.

The mechanism triggering SWT bursts is still unclear. One possible scenario proposed by Boirin et al. (2007), is that the observed delay is caused by a waiting point in the chain of nuclear reactions. A decay reaction with a half-life time similar to the short recurrence time observed, (like the isotope $^{13}\rm N$ with a half-life time of 9.97 min) would stall nuclear burning, and provide the observed waiting times in SWT events. New theoretical models of hydrogen accreting neutron stars are required to reveal the unknown mechanism that halts the H burning. Another possibility could be that the secondary bursts in SWT events result from the ignition of an unburned layer above the ignition depth, which is (rotationally induced) mixed down after the primary burst, down to the depth where a thermonuclear runaway occurs. The rotationally induced mixing may explain recurrence times of $\sim$\,10 minutes (Keek et al. 2010 and references therein). This scenario can explain many of the observed SWT properties, but has not been reproduced by multi-zone stellar evolution models (see Keek et al. 2010 and references therein).

It is interesting to note that our bursts from GS 0836--429 show similar exponential decays ($\sim$ 20 sec), durations ($\sim$ 100 sec), and values of alpha ($\sim$ 50) as the bursts from GS 1826--24 (see, e.g., Galloway et al. 2008), so that the thermonuclear process and fuel composition must be comparable. Keek et al. (2010) reported that neutron stars with spin frequencies $>$ 500 Hz tend to show SWT bursts, so that in a fast rotating neutron star the fresh fuel could mix faster with the ash layers and trigger a secondary burst in time scales of minutes. This would suggest that GS 1826--24 contains a slow rotating neutron star while GS 0836--429 should exhibit a fast spinning neutron star. However, neither for GS 1826--24 nor for GS 0836--429 are neutron-star spin estimates known\footnote{ A tentative high-frequency burst oscillation at 611 Hz from GS 1826--24 was reported by Thompson et al. 2005, but not confirmed, see Watts 2012 for a discussion.}, which would enforce this suggestion.

\begin{acknowledgements}
This work is based on observations with {\it INTEGRAL}, an ESA project with instruments and science data centre funded by ESA member states (especially the PI countries: Denmark, France, Germany, Italy, Switzerland, Spain) and with the participation of Russia and the USA. We are grateful to J\'er\^ome Chenevez for the conversion factor of JEM-X Crab count rate to flux units. 
\end{acknowledgements}

\begin{appendix}
\section{Type I X-ray bursts spectral and light curve parameters}

Various burst parameters are provided in Table~\ref{table:longtables}, including the pointing ID, the start of the burst and the off-axis angle. At high off-axis angles ($> 4^{\circ}$) the data quality is not sufficient enough to extract a time integrated spectrum; for these bursts not all spectral parameters are provided. The same applies to the weak and short bursts such as the SWT bursts. In particular the SWT burst at MJD 52982.65917, that occurs at the end of the pointing; it was not possible to fit the exponential decay in that light curve. Bursts previously detected by Chelovekov et al.\ (2005) are indicated in the MJD column with an asterisk. Parameters extracted from the light curve fitting are also presented in the table, such as the burst recurrence time, the rise time, the e-folding exponential decay time and the peak count rate. The burst temperatures and burst fluences that were obtained from the spectral analysis are also shown. Finally, a measure of the persistent luminosity of the source when the Type I X-ray bursts occurred is given. Note that in various cases there were observation gaps between two consecutive bursts.

%%Font size:
\footnotesize{
%\begin{sidewaystable}
	\longtab{1}{
	 \begin{landscape}
		\begin{longtable}{cccccccccc}
		\caption{\label{table:longtables} Table with parameters of the bursts obtained after the analysis of the light curves and spectra. In the 3rd column we mark the bursts which are also present in the sample from Chelovekov et al. (2005). The last column shows the persistent luminosity of the source at a distance of 9.2\,kpc in the energy range of 5 -- 200 keV.} \\
		
		\hline\hline
		Pointing & Off-axis ($^{\circ}$) & Onset (MJD)  & $\Delta\,t$ (hr) & $\tau$ (s) & Duration (s) & Peak (cts\,s$^{-1}$) & $E_{b}$ ($10^{-7}$\,erg\,cm$^{-2}$) & kT (keV) & $L_{x}$ ($10^{37}$\,erg\,s$^{-1}$) \\ 
		\hline
		\endfirsthead
		\caption{Continued from previous page} \\
		\hline\hline
		Pointing & Off-axis ($^{\circ}$) & Onset (MJD)  & $\Delta\,t$ (hr) & $\tau$ (s) & Duration (s) & Peak (cts\,s$^{-1}$) & $E_{b}$ ($10^{-7}$\,erg\,cm$^{-2}$) & kT (keV) & $L_{x}$ ($10^{37}$\,erg\,s$^{-1}$) \\ 
		\hline
		\endhead
		\hline
		\endfoot
		13700150010.001 & 3.2 & 52970.59361 &  ---  & 19.7$\pm$0.6 & 50 & 55$\pm$14 & 2.9$_{-0.7}^{+0.7}$ & 1.8$_{-0.3}^{+0.4}$ & --- \\ 
		13700240010.001 & 2.4 & 52970.77731 &  4.41 & 32.0$\pm$1.2 & 85 & 39$\pm$8  & 2.5$_{-0.8}^{+1.9}$ & 1.3$_{-0.3}^{+0.4}$ & 1.64$_{-0.22}^{+0.23}$ \\ 
		13700280010.001 & 4.0 & 52970.88039 &  2.47 & 9.8$\pm$0.2 & 30 & 81$\pm$17 & 1.3$_{-0.5}^{+0.8}$ & 1.5$_{-0.4}^{+0.5}$ & 1.82$_{-0.22}^{+0.21}$ \\ 
		13700330010.001 & 2.2 & 52970.97652 &  2.31 & 34.9$\pm$1.0 & 80 & 47$\pm$9  & 2.9$_{-0.6}^{+0.6}$ & 2.4$_{-0.4}^{+0.5}$ & --- \\ 
		13700370010.001 & 4.1 & 52971.06346 &  2.08 & 47.2$\pm$1.3 & 94 & 43$\pm$13 & --- & --- & --- \\ 
		13700750010.001 & 2.1 & 52971.95467$^{*}$ & 21.39 & 18.8$\pm$0.5 & 48 & 61$\pm$9  & 2.4$_{-0.4}^{+0.5}$ & 1.9$_{-0.3}^{+0.3}$ & 1.87$_{-0.14}^{+0.15}$ \\ 
		13700800010.001 & 0.8 & 52972.05397$^{*}$ &  2.38 & 33.3$\pm$1.0 & 88 & 47$\pm$7  & 2.4$_{-0.5}^{+0.5}$ & 1.7$_{-0.3}^{+0.3}$ & 1.96$_{-0.33}^{+0.36}$ \\ 
		13700800010.001 & 0.8 & 52972.06034 &  0.15 & 11.5$\pm$0.8 & 34 & 25$\pm$6  & --- & --- & 1.96$_{-0.33}^{+0.36}$ \\ 
		13700840010.001 & 2.3 & 52972.15864$^{*}$ &  2.36 & 20.5$\pm$0.5 & 52 & 59$\pm$9  & 1.1$_{-0.2}^{+0.3}$ & 1.8$_{-0.3}^{+0.4}$ & 2.00$_{-0.21}^{+0.21}$ \\ 
		13700910010.001 & 3.8 & 52972.32276 &  3.94 & 12.4$\pm$0.3 & 35 & 70$\pm$15 & 2.0$_{-0.5}^{+0.6}$ & 2.4$_{-0.6}^{+0.8}$ & --- \\ 
		13800030010.001 & 2.2 & 52973.40368$^{*}$ & 25.94 & 25.7$\pm$0.7 & 62 & 54$\pm$10 & 2.2$_{-0.4}^{+0.4}$ & 1.8$_{-0.3}^{+0.3}$ & 1.77$_{-0.21}^{+0.22}$ \\ 
		13800080010.001 & 4.0 & 52973.50531 &  2.44 & 7.3$\pm$0.2 & 25 & 64$\pm$16 & 3.7$_{-1.0}^{+1.0}$ & 2.1$_{-0.5}^{+0.6}$ & 1.77$_{-0.20}^{+0.21}$ \\ 
		13800200010.001 & 3.8 & 52973.77511 &  6.48 & 20.3$\pm$0.5 & 51 & 52$\pm$14 & 2.6$_{-0.9}^{+2.4}$ & 1.4$_{-0.4}^{+0.5}$ & 1.91$_{-0.13}^{+0.14}$ \\ 
		13800410010.001 & 3.7 & 52974.24485 & 11.28 & 12.7$\pm$0.4 & 36 & 66$\pm$14 & 2.5$_{-1.0}^{+2.7}$ & 1.3$_{-0.4}^{+0.4}$ & --- \\ 
		13800510010.001 & 2.2 & 52974.45362 &  5.01 & 24.9$\pm$1.9 & 60 & 23$\pm$7  & --- & --- & --- \\ 
		13800590010.001 & 2.2 & 52974.63962$^{*}$ &  4.47 & 10.1$\pm$0.3 & 31 & 57$\pm$9  & 1.3$_{-0.4}^{+0.4}$ & 2.0$_{-0.5}^{+0.6}$ & 1.98$_{-0.15}^{+0.15}$ \\ 
		13800620010.001 & 2.2 & 52974.73375$^{*}$ &  2.26 & 13.7$\pm$0.4 & 27 & 67$\pm$9  & 1.9$_{-0.4}^{+0.4}$ & 2.1$_{-0.5}^{+0.6}$ & 1.97$_{-0.22}^{+0.23}$ \\ 
		13800670010.001 & 3.7 & 52974.84099 &  2.58 & 8.5$\pm$0.5 & 38 & 35$\pm$10 & --- & --- & --- \\ 
		13800710010.001 & 5.2 & 52974.93333 &  2.22 & 30.0$\pm$0.6 & 92 & 70$\pm$29 & --- & --- & --- \\ 
		13800900010.001 & 3.2 & 52975.37217$^{*}$ & 10.53 & 19.1$\pm$0.4 & 49 & 66$\pm$12 & 1.8$_{-0.5}^{+1.0}$ & 1.5$_{-0.4}^{+0.5}$ & --- \\ 
		13800910010.001 & 3.7 & 52975.38611$^{*}$ &  0.33 & 17.3$\pm$0.8 & 45 & 34$\pm$10 & 0.9$_{-0.4}^{+0.8}$ & 1.3$_{-0.6}^{+1.0}$ & --- \\ 
		13800980010.001 & 2.4 & 52975.55933$^{*}$ &  4.16 & 11.2$\pm$1.1 & 33 & 59$\pm$9  & 2.5$_{-0.4}^{+0.4}$ & 2.2$_{-0.3}^{+0.4}$ & 1.85$_{-0.23}^{+0.24}$ \\ 
		13900230010.001 & 3.8 & 52977.08360 & 36.58 & 12.0$\pm$0.6 & 35 & 32$\pm$11 & 1.1$_{-0.6}^{+0.6}$ & 1.7$_{-0.5}^{+0.6}$ & 1.77$_{-0.19}^{+0.20}$ \\ 
		13900320010.001 & 2.4 & 52977.29309$^{*}$ &  5.03 & 12.5$\pm$0.4 & 46 & 65$\pm$10 & 4.0$_{-1.1}^{+2.0}$ & 1.4$_{-0.3}^{+0.4}$ & 1.88$_{-0.23}^{+0.24}$ \\ 
		13900360010.001 & 4.0 & 52977.38241 &  2.14 & 23.0$\pm$0.8 & 57 & 44$\pm$14 & 2.1$_{-0.6}^{+0.9}$ & 2.1$_{-0.3}^{+0.3}$ & --- \\ 
		13900400010.001 & 0.8 & 52977.46563$^{*}$ &  2.00 & 12.0$\pm$0.3 & 35 & 75$\pm$9  & 1.8$_{-0.3}^{+0.3}$ & 2.2$_{-0.3}^{+0.3}$ & 1.90$_{-0.32}^{+0.35}$ \\ 
		13900450010.001 & 4.1 & 52977.57044 &  2.51 & 23.2$\pm$0.5 & 46 & 66$\pm$16 & 2.9$_{-1.3}^{+7.1}$ & 1.2$_{-0.5}^{+0.5}$ & 1.61$_{-0.18}^{+0.19}$ \\ 
		13900730010.001 & 3.7 & 52978.22162 & 15.63 & 11.8$\pm$0.3 & 24 & 60$\pm$25 & 1.3$_{-0.5}^{+0.6}$ & 1.5$_{-0.4}^{+0.4}$ & --- \\ 
		13900770010.001 & 4.1 & 52978.31238 &  2.18 & 37.8$\pm$1.2 & 76 & 40$\pm$13 & 2.9$_{-0.8}^{+0.9}$ & 1.9$_{-0.5}^{+0.6}$ & --- \\ 
		13900800010.001 & 2.4 & 52978.39566 &  2.00 & 11.2$\pm$0.4 & 33 & 54$\pm$9  & 2.5$_{-0.7}^{+0.8}$ & 1.9$_{-0.5}^{+0.9}$ & 1.61$_{-0.23}^{+0.24}$ \\ 
		13900890010.001 & 2.2 & 52978.58350 &  4.51 & 13.3$\pm$0.2 & 37 & 90$\pm$12 & 3.4$_{-0.6}^{+0.7}$ & 2.0$_{-0.4}^{+0.4}$ & 1.70$_{-0.14}^{+0.15}$ \\ 
		13900930010.001 & 4.1 & 52978.68097 &  2.34 & 17.8$\pm$0.4 & 46 & 68$\pm$17 & 1.7$_{-0.5}^{+0.5}$ & 2.2$_{-0.5}^{+0.6}$ & 1.72$_{-0.20}^{+0.21}$ \\ 
		13900980010.001 & 3.2 & 52978.79877$^{*}$ &  2.83 & 25.1$\pm$0.5 & 50 & 54$\pm$11 & 1.5$_{-0.4}^{+0.4}$ & 1.8$_{-0.4}^{+0.5}$ & 1.68$_{-0.56}^{+0.62}$ \\ 
		14000230010.001 & 2.3 & 52979.84076$^{*}$ & 25.01 & 15.8$\pm$0.3 & 32 & 84$\pm$10 & 5.8$_{-0.8}^{+0.9}$ & 1.7$_{-0.2}^{+0.2}$ & 1.81$_{-0.38}^{+0.41}$ \\ 
		14000270010.001 & 0.8 & 52979.93076$^{*}$ &  2.16 & 26.9$\pm$2.3 & 64 & 48$\pm$7  & 1.4$_{-0.3}^{+0.3}$ & 2.2$_{-0.4}^{+0.4}$ & 1.67$_{-0.20}^{+0.21}$ \\ 
		14000320010.001 & 2.1 & 52980.03770$^{*}$ &  2.57 & 16.0$\pm$0.4 & 43 & 62$\pm$9  & 2.4$_{-0.4}^{+0.4}$ & 2.1$_{-0.3}^{+0.4}$ & --- \\ 
		14000360010.001 & 1.2 & 52980.12437$^{*}$ &  2.08 & 16.1$\pm$0.4 & 43 & 57$\pm$8  & 2.7$_{-0.6}^{+0.9}$ & 1.4$_{-0.3}^{+0.3}$ & 1.55$_{-0.18}^{+0.19}$ \\ 
		14000640010.001 & 3.8 & 52980.76314 & 15.33 & 28.5$\pm$1.2 & 68 & 37$\pm$12 & 2.8$_{-0.5}^{+0.5}$ & 1.8$_{-0.3}^{+0.3}$ & --- \\ 
		14000720010.001 & 1.2 & 52980.95392$^{*}$ &  4.58 & 16.2$\pm$0.4 & 32 & 61$\pm$8  & 2.3$_{-0.3}^{+0.3}$ & 2.1$_{-0.2}^{+0.2}$ & 1.87$_{-0.34}^{+0.36}$ \\ 
		14000760010.001 & 2.1 & 52981.04882$^{*}$ &  2.28 & 17.6$\pm$0.6 & 46 & 44$\pm$8  & 1.5$_{-0.5}^{+1.1}$ & 1.4$_{-0.4}^{+0.6}$ & 1.47$_{-0.18}^{+0.19}$ \\ 
		14000810010.001	& 0.8 & 52981.14644$^{*}$ & 2.34 & 25.9$\pm$0.7 & 62 & 57$\pm$8	 & 1.1$_{-0.2}^{+0.2}$ & 1.9$_{-0.3}^{+0.4}$ & 1.59$_{-0.32}^{+0.34}$ \\
		14000850010.001 & 2.3 & 52981.24207$^{*}$ &  2.30 & 33.2$\pm$0.8 & 77 & 60$\pm$9  & 1.4$_{-0.3}^{+0.3}$ & 2.1$_{-0.4}^{+0.6}$ & 1.52$_{-0.18}^{+0.19}$ \\ 
		14000890010.001 & 5.0 & 52981.32780 &  2.06 & 12.6$\pm$0.3 & 44 & 67$\pm$25 & --- & --- & --- \\ 
		14100060010.001 & 1.2 & 52982.42880$^{*}$ & 26.42 & 24.6$\pm$0.7 & 49 & 48$\pm$7  & 2.2$_{-0.4}^{+0.6}$ & 1.5$_{-0.3}^{+0.3}$ & 1.80$_{-0.35}^{+0.37}$ \\ 
		14100110010.001 & 4.0 & 52982.53187 & 2.47 & 7.6$\pm$0.2 & 26 & 77$\pm$16 & 2.8$_{-0.6}^{+0.6}$ & 2.0$_{-0.4}^{+0.4}$  & 1.45$_{-0.16}^{+0.17}$ \\ 
		14100160010.001 & 2.3 & 52982.65130 &  2.87 & 18.2$\pm$0.7 & 57 & 37$\pm$8  & 1.6$_{-0.5}^{+0.6}$ & 1.5$_{-0.4}^{+0.4}$ & 1.63$_{-0.26}^{+0.28}$ \\ 
		14100160010.001 & 2.3 & 52982.65917 &  0.19 & --- & 60 & 23$\pm$7  & --- & --- & 1.63$_{-0.26}^{+0.28}$\\ 
		14100200010.001 & 4.1 & 52982.74237 &  2.00 & 21.1$\pm$0.8 & 53 & 42$\pm$14 & 1.5$_{-0.7}^{+0.7}$ & 2.0$_{-0.5}^{+0.9}$ & 1.45$_{-0.19}^{+0.20}$ \\ 
		14100240010.001 & 3.7 & 52982.82781 &  2.05 & 25.8$\pm$0.7 & 73 & 55$\pm$11 & 3.7$_{-0.7}^{+0.7}$ & 2.4$_{-0.4}^{+0.5}$ & --- \\ 
		14100460010.001 & 3.8 & 52983.32016 & 11.82 & 30.1$\pm$1.5 & 71 & 42$\pm$13 & 2.1$_{-0.6}^{+0.6}$ & 2.2$_{-0.6}^{+0.6}$ & --- \\ 
		14100550010.001 & 2.4 & 52983.50750$^{*}$ &  4.49 & 23.6$\pm$0.6 & 58 & 44$\pm$9  & 1.9$_{-0.6}^{+1.0}$ & 1.5$_{-0.4}^{+0.4}$ & 1.55$_{-0.22}^{+0.23}$ \\ 
		14100550010.001 & 2.4 & 52983.51408 &  0.16 &  15.5$\pm$1.1 &  40 & 19.4$\pm$6.4 & --- & --- & 1.55$_{-0.22}^{+0.23}$ \\   
		14100590010.001 & 4.0 & 52983.60479 &  2.18 & 10.9$\pm$0.3 & 32 & 57$\pm$15 & 2.6$_{-0.5}^{+0.6}$ & 1.7$_{-0.2}^{+0.3}$ & 1.49$_{-0.18}^{+0.19}$ \\ 
		14100630010.001 & 0.8 & 52983.71024$^{*}$ &  2.53 & 16.5$\pm$0.5 & 44 & 56$\pm$8  & 1.5$_{-0.3}^{+0.3}$ & 2.4$_{-0.4}^{+0.5}$ & 1.42$_{-0.37}^{+0.43}$ \\ 
		14100680010.001 & 4.1 & 52983.82023 &  2.64 & 14.6$\pm$0.2 & 40 &101$\pm$18 & 3.2$_{-0.7}^{+0.7}$ & 1.9$_{-0.3}^{+0.4}$ & 1.36$_{-0.16}^{+0.17}$ \\ 
		14100950010.001 & 3.2 & 52984.44225$^{*}$ & 14.93 & 8.7$\pm$0.2 & 28 & 88$\pm$13 & 3.0$_{-0.5}^{+0.5}$ & 2.2$_{-0.3}^{+0.3}$ & 1.69$_{-0.44}^{+0.48}$ \\ 
		14101000010.001 & 4.0 & 52984.54426 &  2.45 & 24.3$\pm$1.0 & 59 & 44$\pm$14 & 1.7$_{-0.9}^{+3.0}$ & 1.4$_{-0.6}^{+0.9}$ & 2.15$_{-1.02}^{+1.40}$ \\ 
		14101040010.001 & 2.3 & 52984.63308 &  2.13 & 16.4$\pm$0.5 & 43 & 49$\pm$9  & 1.5$_{-0.4}^{+0.4}$ & 1.9$_{-0.4}^{+0.5}$ & 1.40$_{-0.23}^{+0.24}$ \\ 
		14101090010.001 & 4.1 & 52984.74813 &  2.76 & 14.1$\pm$0.4 & 39 & 63$\pm$16 & 3.4$_{-0.7}^{+0.7}$ & 2.2$_{-0.4}^{+0.4}$ & 1.26$_{-0.21}^{+0.22}$ \\ 
		14900030010.001 & 2.2 & 53006.42482 & ---   & 23.9$\pm$0.8 & 58 & 44$\pm$9  & 2.1$_{-0.6}^{+0.6}$ & 2.0$_{-0.5}^{+0.6}$ & 1.49$_{-0.16}^{+0.16}$ \\ 
		18600820010.001 & 1.8 & 53119.17252 & ---   & 13.2$\pm$0.3 & 47 & 68$\pm$9  & 2.0$_{-0.4}^{+0.4}$ & 2.0$_{-0.3}^{+0.3}$ & --- \\ 
		\end{longtable}
	\label{table:longtables}
	\end{landscape}
	}
%\end{sidewaystable}
} %close footnotesize

\end{appendix}

\end{document}